\documentclass[12pt]{elsarticle}
\usepackage[utf8]{inputenc}
\usepackage{graphicx}
\usepackage{subfigure}
\usepackage{array}
\usepackage{verbatim}

\usepackage{hyperref}

\usepackage[detect-all]{siunitx}

\usepackage{amssymb}
\usepackage{amsthm}
\usepackage{url}
\usepackage{amsmath}
\usepackage{subfigure}
\usepackage{textcomp}
\usepackage{xcolor}

\graphicspath{{figs/}}

\journal{Biosystems}

\begin{document}

\begin{frontmatter}
\title{Propagation of electrical signals by fungi}

\author[1,2]{Richard Mayne}
\author[1,3]{Nic Roberts}
\author[1]{Neil Phillips}
\author[4]{Roshan Weerasekera}
\author[1]{Andrew Adamatzky} 

\address[1]{Unconventional Computing Laboratory, University of the West of England, Bristol, UK}
\address[2]{Mayne Bio Analytics Ltd., Cinderford, UK}
\address[3]{Advanced Machinery and Productivity Institute, University of Huddersfield, Huddersfield, UK}
\address[4]{School of Computer Science, Electrical and Electronic Engineering and Engineering Maths, University of Bristol, Bristol, UK}


\begin{abstract}
Living fungal mycelium networks are proven to have properties of memristors, capacitors and various sensors. To further progress our designs in fungal electronics we need to evaluate how electrical signals can be propagated through mycelium networks. We investigate the ability of mycelium-bound composites to convey electrical signals, thereby enabling the transmission of frequency-modulated information through mycelium networks. Mycelia were found to reliably transfer signals with a recoverable frequency comparable to the input, in the \SIrange{100}{10000} {\hertz} frequency range. Mycelial adaptive responses, such as tissue repair, may result in fragile connections, however.  While the mean amplitude of output signals was not reproducible among replicate experiments exposed to the same input frequency, the variance across groups was highly consistent. Our work is supported by NARX modelling through which an approximate transfer function was derived. These findings advance the state of the art of using mycelium-bound composites in analogue electronics and unconventional computing.

\end{abstract}


\end{frontmatter}

\section{Introduction}

Manufacturing electronic devices makes negative environmental impacts especially when a disposal of electronic devices is concerned~\cite{cenci2021eco}. An alternative eco-friendly approach could be to use of organic living materials to build electronic devices~\cite{han2020advanced,li2020biodegradable,wu2021biodegradable}. Living computing substrates might create novel properties impossible to replicate with silicon, expanding the world of computing and electronics~\cite{lee2017toward,chang2017circuits,van2018organic,friederich2019toward}. 
The sister field of organic electronics is unconventional computing, which uncovers novel principles of efficient information processing in physical, chemical, biological systems, to develop novel computing substrates, algorithms, and architectures~\cite{adamatzky2016advances,adamatzky2021thoughts}.

Whilst a substantial number of organic electronic devices offer a high degree of flexibility in the application domain, few prototypes show stability and biocompatibiliy~\cite{feron2018organic}. Electronic devices made from living fungi might offer a feasible opportunity to design and prototype self-growing living circuits. In \cite{adamatzky2022fungal} we introduced fungal electronics --- a family of living electronic devices made of mycelium bound composites or pure mycelium. Fungal electronic devices are capable of changing their impedance and generating spikes of electrical potential in response to external control parameters. Fungal electronics can be embedded into fungal materials and wearables or used as stand alone sensing and computing devices. In experimental laboratory conditions we demonstrated that fungi can be used as memristors~\cite{beasley2022mem}, photosensors~\cite{beasley2020fungal,adamatzky2021towards}, chemical sensors~\cite{dehshibi2021stimulating,adamatzky2021reactive}, humidity sensors~\cite{phillips2023electrical}, and tactile sensors~\cite{adamatzky2021towards}. To make a functional fungal circuits one must connect several fungal electronic devices. Thus, in present paper we analyse how suitable mycelium networks could be for propagating electric signals.

The elucidation of signalling mechanisms within and between fungi is furthermore significant to numerous fields of study. It furthers our efforts to define data transfer between living organisms as a fundamentally algorithmic process and helps us understand the factors that define the dynamics of the natural environment, such as host-fungal interactions in both synergistic and pathogenic contexts.

Using basidiomycete \emph{Ganoderma lucidum} as a model organism, we examine the potential for information transmission via mycelium networks. In particular, whether it is possible to recover precise frequency information from an input signal that has traversed an inoculated substrate, allowing frequency modulated information to be transmitted through mycelial structures. 

Through data-driven modelling, we extend our analysis and generalise the dynamics and signal transfer function therein. Our works are supported through both laboratory experimental measurement and a complimentary nonlinear autoregressive exogenous model (NARX) model. We conclude by discussing the applications of electrical signal transfer in fungi.

\section{Methods}
\subsection{Mycelium cultivation}

\emph{Ganoderma lucidum} (strain M9724, Mycelia, BE) was cultivated on a rye seed and millet grain medium with a 1\% spawn inoculation rate. Incubation was conducted for approximately 18 days at room temperature (\SI{23}{\degreeCelsius}), until a contiguous layer of mycelium was visible on both the upper and lower surfaces of the medium; initial experimentation demonstrated that the interior surfaces of the medium were well-colonised under these conditions. When mycelia were not used immediately for experiments, they were sealed in plastic bags and refrigerated until use.

\subsection{Electrical stimulation and measurement}

\begin{figure}[h]
    \centering
    \subfigure[]{\includegraphics[width=0.95\textwidth]{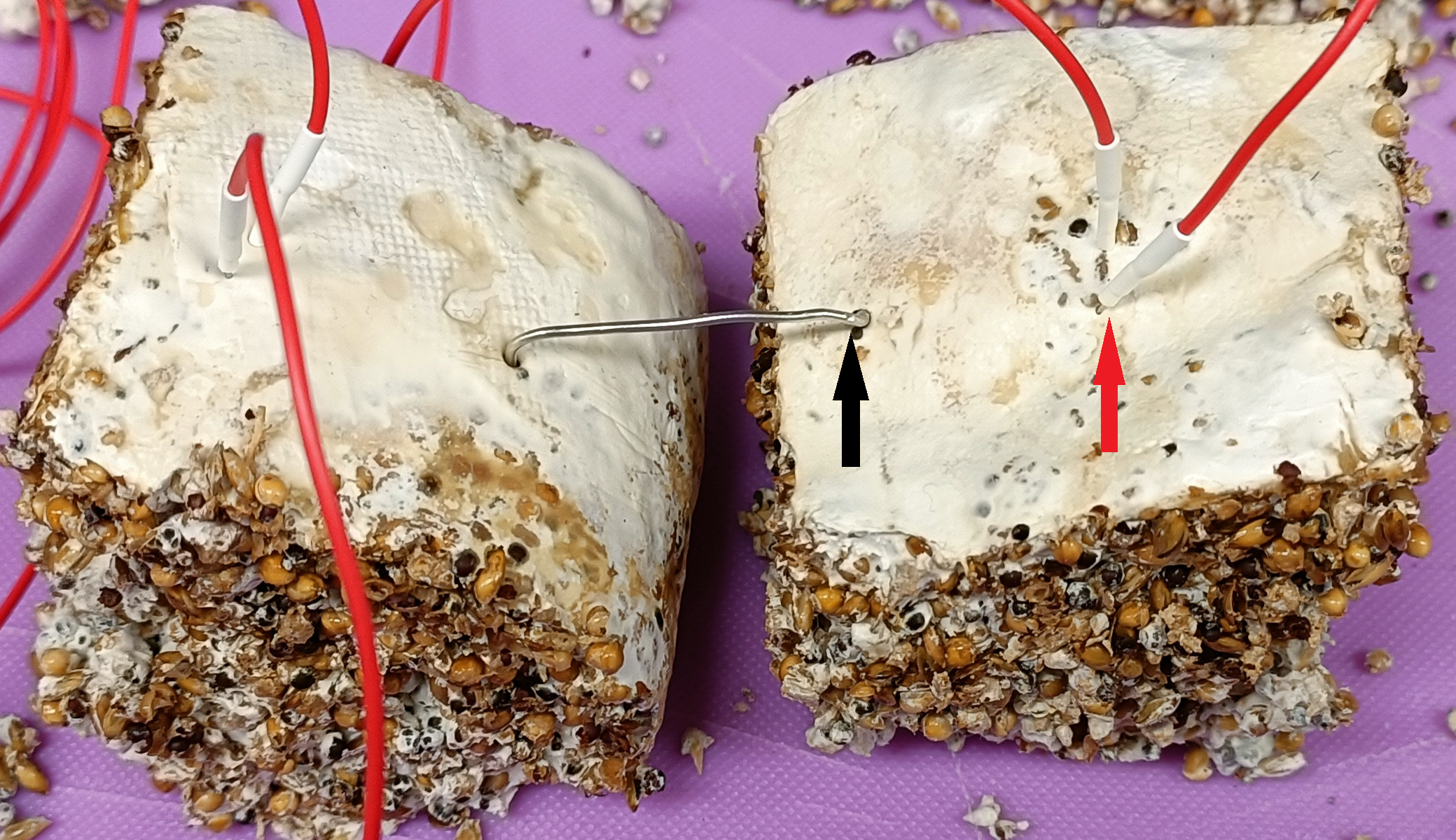}}
    \subfigure[]{\includegraphics[width=0.95\textwidth]{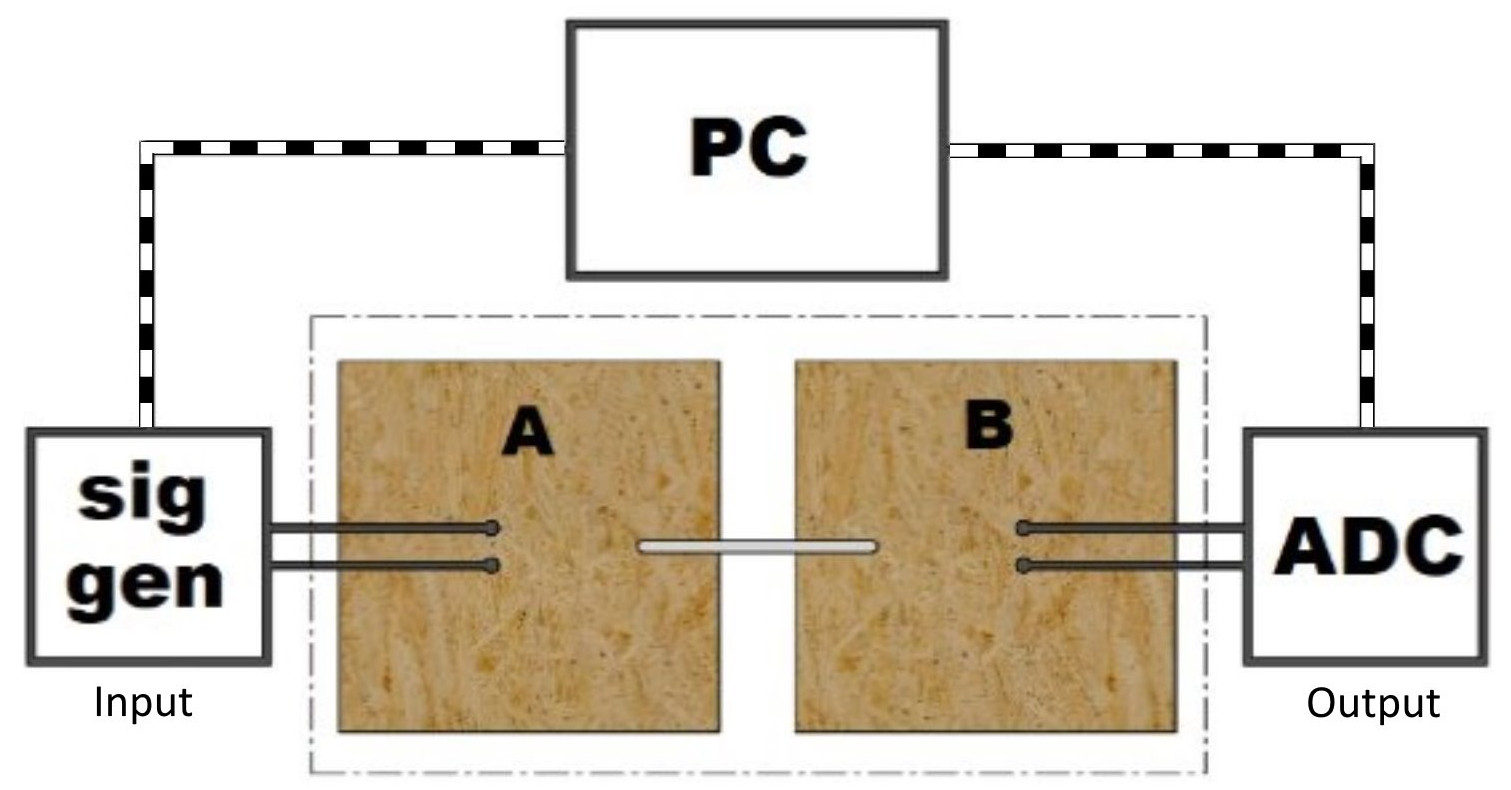}}
    \caption{(a) Plan view of two mycelium blocks connected together by platinum wire (black arrow), showing electrode connections (red arrow). White patches are mycelium. (b) Schematic of computer (PC) controlled input from signal generator (`sig gen') and recording of the output via analogue to digital converter (ADC). }
    \label{fig:Blocks}
\end{figure}

For each set of measurements, two blocks `A' and `B', each measuring $\sim$$50~\times 50\times~$\SI{50} {\mm} were cut from a substrate that had been well colonised by \emph{G. lucidum}. The blocks were separated by \SI{10} {\mm} and electrically connected by a by a \SI{1} {\mm} diameter platinum wire of length \SI{40} {\mm}, see  Fig.~\ref{fig:Blocks}.

Control experiments consisted of an identical setup with the exception that colonised blocks were replaced with blocks of compressed, clean culture medium that had been soaked in excess tap water for 30 minutes prior to beginning experiments. An electrically insulating resin board (resistance $>$\SI{200}{\mega\ohm\per\milli\meter}) was used to under the experimental apparatus.

The `input' and `output' connections were Pt/Ir needle electrodes (Spes Medica S.r.l., IT) separated by \SI{5} {\mm}. Electrodes were positioned \SI{20} {\mm} from the platinum wire in in both blocks. After the needle electrodes were inserted, the colonised substrate was left for  $\sim$\SI{24} {\hour} to allow the mycelium and electrodes to form electrical connections.

Input signals were {$\pm$}\SI{5} {\volt} square waves generated by AD9833 signal generator modules (Analog Devices Inc., USA) and controlled by a PC via an Arduino Mega (Elegoo Inc., China) analog-to-digital converter (ADC). In order to directly compare measured signals with stimulation, the ADC was also used to measure the input signal in synchronisation with block B measurements (see Sec. \ref{sec-analysis}). The frequencies employed to stimulate block A ranged \SIrange{100}{1000} {\hertz} (\SI{100} {\hertz} intervals) and \SIrange{1}{10} {\kilo\hertz} (\SI{1} {\kilo\hertz} intervals), with each frequency being applied for \SI{60} {\second}, for a total of 19 frequency tests (n = 28 for each frequency).

Both input and output signals were sampled at \SI{50} {\kilo\hertz} using a 16-bit analogue to digital converter NI-USB-6210 (National Instruments, Austin, USA).

\subsection{Analysis}
\label{sec-analysis}

All data analysis was conducted using Python 3.10 and graphing was done via Matplotlib 3.7 \cite{matplotlib2007}. Numerical work consisted of:

\begin{enumerate}
    \item[a] Testing the hypothesis that output signals were derived from input signals (as opposed to no signal or the fundamental signalling events observed in Ref. \cite{dehshibi2021electrical}).
    \item[b] Quantifying how input signals became altered by transmission through two discrete mycelial networks. 
\end{enumerate}

Statistical analysis was done for criterion (a) and modelling was conducted for (b) (see Sec. \ref{sec-modelling}).

Initial analyses used Cross Spectral Density (CSD) in conjunction with Welch's method via SciPy 1.10.1 \cite{scipy2020} to compare the output signal's strongest harmonic components to the input signal's fundamental frequency. All time series were subject to significant autocorrelation, partial autocorrelation, and non-stationarity, as input signals were periodic and output signals were likely to be subject to some adaptive responses from the mycelia.

Further, identification of an output frequency corresponding to an input frequency using this method is only indicative of signal transfer, as estimation of power is unreliable for nonstationary time series. 

The following descriptive statistics were additionally gathered using SciPy, NumPy 1.24.2 \cite{numpy2020}, and StatsModels 0.13.5 \cite{statsmodels2010} libraries:

\begin{enumerate}
    \item Test of specific distribution of each sample, via Anderson-Darling test.
    \item Number of samples with a recoverable frequency of an order divisible by the input frequency, via comparing the highest power frequency (rounded to 2 significant figures) obtained via CSD with the input frequency. Results were collected as percentage of samples which had a recoverable frequency, per input frequency group.
    \item Dominant amplitude, extracted via discrete Fourier transform (DFT). Results collected as theaverage amplitude per input frequency group. Additional tests were conducted to determine whether average values of samples within input frequency categories were equal, which was achieved via Kruskal-Wallis test. 
    \item Stationarity, measured by Augmented Dickey Fuller (ADF) test. Results collected as percentage of non-stationary samples, per input frequency group.
    \item Granger causality to assess whether a causal relationship were likely present between input and output signals \cite{dean2016}, using the Chi-Square distribution. Max lags were determined quantitatively via cross-correlation. Specifically, the test examines whether past values of the input signal have a statistically significant effect on the momentary value of the output signal, taking past output signals as regressors. Data were initially stationarized through differencing input and output signals. Results collected as a percentage of samples in which the null hypothesis were rejected, per each input frequency.
\end{enumerate}

Usefully, statistically significant Granger causality also indicates whether a relationship between two time series may be adequately described with a linear component, results of which were used to inform approaches to answer (b). 

\subsection{Modelling}
\label{sec-modelling}
Modelling was performed using Python 3.10 and SysIdentPy 0.2.1 \cite{Lacerda2020sysidentpy}. The purpose of modelling was (a) to quantify the relationship between input and output signals for a given treatment via approximation of transfer function; (b) comment on variability of transfer functions between organisms. 

Nonlinear AutoRegressive Exogeous (NARX) modelling was chosen as our approach for its demonstrable value in predicting time series values given past values with respect to a delay parameter (lags), rationale being that autoregressive time series forecasting may be used as an indicator of model fit \cite{Lacerda2020sysidentpy}. Even in instances where observed behaviour are linear, NARX models identify and, if necessary, separate nonlinear phenomena, through a model structure selection process which identifies components of input signals which are useful in forecasting future values and eliminating the non-useful \cite{araujo2019frols}. 

Specifically, we employ Forward Regression Orthogonal Least Squares algorithm (FROLS) as a method of selecting the most useful regressors from our system (Eq. \ref{eq-frols}) \cite{chen1989frols}. This step is necessary to reduce overfitting and was selected as part of preliminary data analysis (data not shown). 

\begin{equation}
\label{eq-frols}
    y_k= F^\ell[y_{k-1}, \dotsc, y_{k-n_y},x_{k-d}, x_{k-d-1}, \dotsc, x_{k-d-n_x}, e_{k-1}, \dotsc, e_{k-n_e}] + e_k
\end{equation}

Where $n_y, n_x, n_e\in \mathbb{N}$ are the maximum lags for the system output, input and input noise regressors, respectively; $x_k \in \mathbb{R}^{n_x}$ is the system input and $y_k \in \mathbb{R}^{n_y}$ is the system output at discrete time $k \in \mathbb{N}^n$ and $e_k \in \mathbb{R}^{n_e}$ is uncertainties and possible noise in discrete time. $\mathcal{F}^\ell$ is the function of input and output regressors with nonlinearity degree $\ell \in \mathbb{N}$, and $d$ is a fixed time delay \cite{Lacerda2020sysidentpy}.

Modelling experiments focused on only one input treatment, arbitrarily chosen as 900~Hz, as the transfer function is unique to the input. As will be described below, this process consisted of building a candidate model through single in-single out methodology (training), then running predictions on time series not seen during the training process (validation).  

Initially, parameter estimation was conducted by doing a grid search through a computationally feasible hyperparameter search space, candidates for which were basis function (polynomial, Fourier), basis function degree, maximum number of terms (regressors) and use of extended least squares regression. Maximum lags were determined quantitatively for each sample through autocorrelation functions, see Sec \ref{sec-analysis}. Models were trained against 22 samples of \SI{100000} time steps, using an 80/20 train/test split and relative square root error (RRSE) as the performance metric. 

Model performance was validated against 6 time series, reporting of which was through mean and variance of the RRSE from these runs, in addition to the estimated transfer function.

\section{Results}
\subsection{General characteristics \& Statistics}
\begin{figure} [htbp!]
   \centering
   \includegraphics[width=\textwidth]{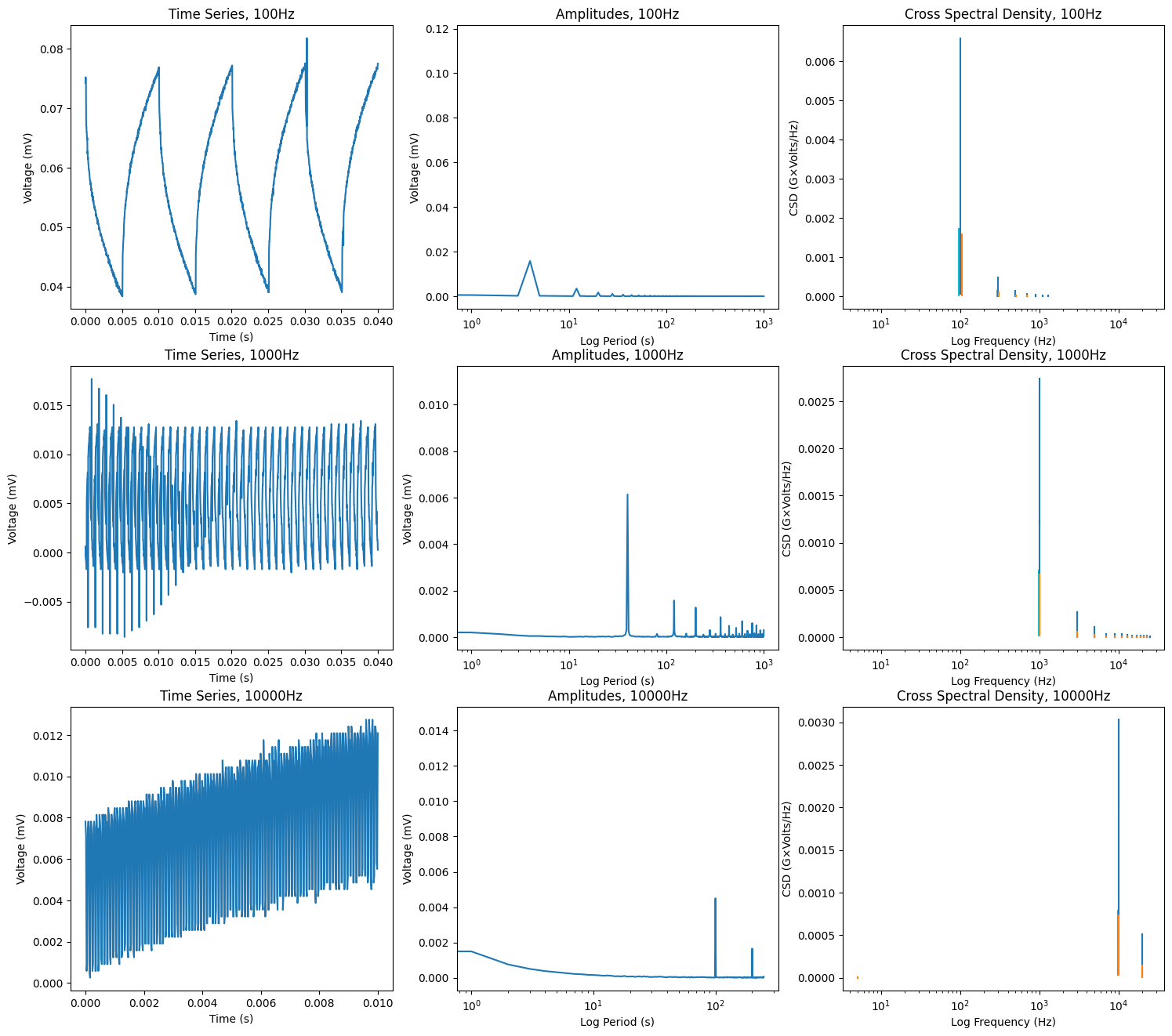}
   \caption{Representative (left) input and output signals, (centre) recovered amplitudes and (right) cross spectral densities from three separate sets of mycelia stimulated with 100~Hz, 1~KHz and 10~KHz square waves.}
   \label{fig-waves}
\end{figure}

Exemplar plots from several input frequencies are shown in Fig.~\ref{fig-waves}. Qualitative interpretation of graphs revealed waves whose frequencies were superficially similar to the input signal were present in the majority of samples. Wave shapes tended towards being closer to sawtooth than square in appearance, but were frequently irregular and occasionally exhibited envelope-like structures (See Fig.~\ref{fig-waves} 100~Hz time series). 

No signals were measured in control experiments, indicating that neither the growth medium or interstitial moisture were capable of conducting the input signals between blocks A and B.

Descriptive statistics in Tab. \ref{tab-desc_stats} illustrate that the majority of output signals across all input frequencies transmit a recoverable frequency comparable to the input frequency. Further, most samples were also non-stationary as indicated by the Augmented Dickey Fuller test, indicating that the mean voltage measurement was likely to change over time, potentially indicating an adaptive response by the mycelia.

\begin{table}[ht]
    \centering
    \begin{tabular}{c|c|l|c|c}
    Input Frequency & PcRF & MA (IQR) & PcADF & PcGC\\
    \hline
    \hline
     100 Hz     & 100.00  & 0.20 (0.17) $\dagger$ & 82.14  & 100.00 \\
     200 Hz     & 100.00  & 0.09 (0.06) $\dagger$ & 75.00  & 96.43 \\
     300 Hz     & 100.00  & 0.07 (0.05) $\dagger$ & 60.71  & 100.00 \\
     400 Hz     & 100.00  & 0.07 (0.06) $\dagger$ & 35.71  & 100.00 \\
     500 Hz     & 100.00  & 0.08 (0.06) $\dagger$ & 35.71  & 85.71 \\
     600 Hz     & 100.00  & 0.09 (0.09) $\dagger$ & 39.29  & 92.86 \\
     700 Hz     & 100.00  & 0.08 (0.09) $\dagger$ & 53.57  & 100.00 \\
     800 Hz     & 96.43   & 0.08 (0.09) $\dagger$ & 60.71  & 100.00 \\
     900 Hz     & 96.43   & 0.08 (0.10) $\dagger$ & 57.14  & 100.00 \\
     1 KHz      & 100.00  & 0.09 (0.10) $\dagger$ & 50.00  & 75.00 \\
     2 KHz      & 100.00  & 0.08 (0.11) $\dagger$ & 100.00  & 75.00 \\
     3 KHz      & 82.14   & 0.08 (0.10) $\dagger$ & 100.00  & 96.43 \\
     4 KHz      & 85.71   & 0.08 (0.11) $\dagger$ & 92.86  & 82.14 \\
     5 KHz      & 100.00  & 0.08 (0.09) $\dagger$ & 71.43  & 96.43 \\
     6 KHz      & 60.71   & 0.07 (0.10) $\dagger$ & 96.43  & 96.43 \\
     7 KHz      & 71.43   & 0.08 (0.11) $\dagger$ & 85.71  & 100.00\\
     8 KHz      & 78.57   & 0.09 (0.09) $\dagger$ & 96.43  & 96.43 \\
     9 KHz      & 46.43   & 0.08 (0.09) $\dagger$ & 75.00  & 100.00 \\
     10 KHz     & 96.43   & 0.08 (0.09) $\dagger$ & 39.29  & 78.57 \\
     \hline 
     Median (IQR)& 100.00 (16.08)  & 0.09 (0.10)  &  71.43 (37.50) & 96.43 (10.72)\\
    \end{tabular}
    \caption{Descriptive statistics on time series. ``PcRF'' Percentage of replicate recordings with recoverable input frequency, percent; ``MA (IQR)'' Median of highest power amplitude with interquartile range, millivolts, $\dagger$: at least one median within within group were significantly different via Kruskal-Wallis test, P $< 0.001$; ``PcADF'' Percentage of replicate recordings with Augmented Dickey Fuller test for stationarity significance at P $<0.05$, percent; ``Median (IQR)'' Median and interquartile range for column. ``PcGC" Percentage of replicate recordings in which Granger Causality hypothesis were rejected, percent.}
    \label{tab-desc_stats}
\end{table} 

The majority of samples had recoverable frequencies that were equal or next-order harmonics of the input signal. The repeatability of recovering these frequencies reduced as input frequency increased. It was not clear from this statistic whether samples that did not have a recoverable frequency were regardless transmitting a periodic signal.

The median amplitude of measured signals within input frequency groups was highly variable, as was evidenced by significance with the Kruskal-Wallis test for every input frequency. There were no significant differences in median amplitude between input frequency groups, however, indicating similar levels of variance. 

Most samples were also demonstrated to Granger Cause. No correlations were observed between samples with/without recoverable frequencies, non-/stationarity or non-/significant Granger Causation. 

\subsection{Modelling}
\begin{figure} [htbp]
    \centering
    \subfigure[]{\includegraphics[width=0.8\textwidth]{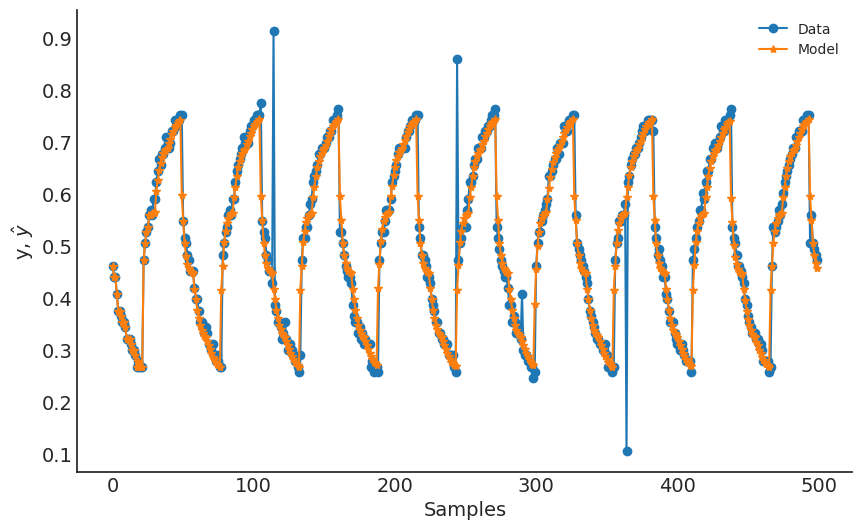}}
    \subfigure[]{\includegraphics[width=0.8\textwidth]{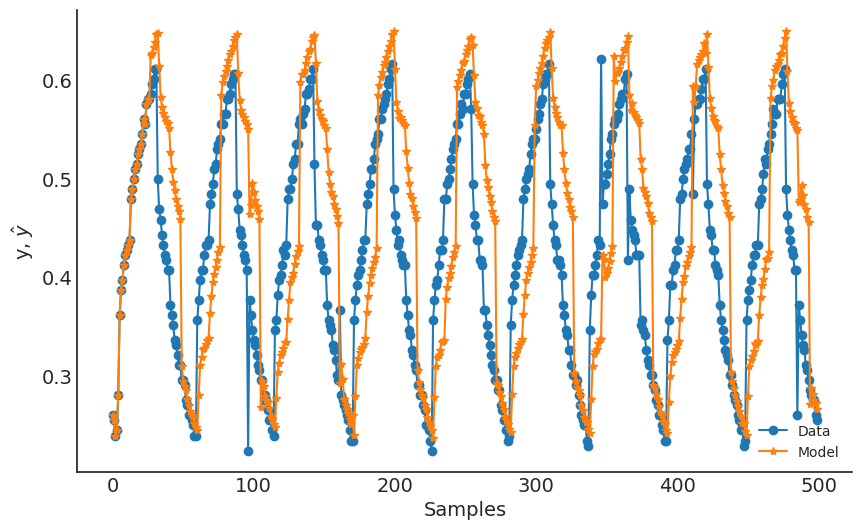}}
    \caption{Representative NARX modelling of input-output transfer function, showing time series data and model predictions. (a) Test data free run simulation, RRSE 0.255. (b) Validation data free run simulation, RRSE 0.748.}
    \label{fig-model}
\end{figure}

Optimal hyperparameters identified in model selection are shown in Tab. \ref{tab-hparams}. Signal transfer was found to be approximated well with a comparatively simple five-term first order polynomial, indicating that the dominant portion of the signal measured was likely derived from the simple input square wave. Eq. \ref{eq-model} shows the approximate transfer function and Fig. \ref{fig-model} shows representative output from both training and validation stages.

\begin{table}[hbtp]
    \centering
    \begin{tabular}{c|c}
    Hyperparameter  &  Value \\
    \hline
    \hline
    Basis function   & Polynomial \\
    Degree           & 1          \\
    N Terms          & 5          \\
    ELSR             & False      \\
    \end{tabular}
    \caption{Hyperparameters for selected model. ELSR: extended least squares regression.}
    \label{tab-hparams}
\end{table}

\begin{multline}
    \label{eq-model}
    \begin{split}
    y(k) = &~ 0.33\pm 0.99 \cdot y(k-1) + 0.21\pm 5.5\times 10^{-4} \cdot y(k-2) \\ & - 0.15\pm 9.2\times 10^{-4} \cdot y(k-27) + 0.20\pm 1.3\times 10^{-3} \cdot y(k-9) \\ & + 0.21\pm 4.2\times 10^{-3}
    \end{split}
\end{multline}

We may observe that the transfer function has wide error margins on the $y(k-1)$ term, corresponding to the single step ahead. Aggregate test set RRSE was 0.255; validation set RRSE was 0.748.

\section{Discussion}

Our primary finding is that mycelia do support modes of electrical communication between them. The majority of samples we measured were found to transmit a signal of a frequency comparable to the input signal. The Granger Causality test, while not an indication of true causality, demonstrates that the input signal is helpful in forecasting the measured signal. We find therefore that these two statistics are highly indicative that the output signal was derived from input signal in the majority of experiments.

In the light of recent work done on electrochemical signalling between discrete organisms, our results indicate that such communication is theoretically possible between mycelia.

There are several reasons for which samples did not exhibit a recoverable frequency, but broadly this occurred when either no signal was transmitted, or when the input signal was modified sufficiently to not be a harmonic of the input signal. Instances for the former count result when no electrical connection was made with the organism --- more explicitly, the surface area of electrode in contact with mycelium was insufficient to conduct a measurable signal. Reasons for poor connections could include adaptive responses from the mycelia, such as tissue damage and repair.  

We observed no relationship between the percentage of samples that had no recoverable frequency, those exhibiting non-stationarity or those with significant Granger Causality. We envisage several scenarios for explaining this phenomenon. Firstly, samples with very weak or entirely absent output signals --- possibly due to a failed electrical connection on one or more of the electrodes --- would be unlikely to have a recoverable frequency or Granger Cause, but would be stationary. Secondly, samples where the input signal were transformed to a non-harmonic frequency would have no recoverable frequency in our measurement scheme, but would potentially Granger Cause and/or be non-stationary. Thirdly, samples with a recoverable frequency but exhibiting unusual wave patterns, potentially due to interference of electrochemical events such as calcium waves \cite{dehshibi2021electrical}, would be unlikely to be stationary and may introduce enough uncertainty to lead to failure to Granger Cause. These observations highlight the inherent variability of living systems and hence the requirement for practical unconventional computing device design to compensate for or mitigate these effect.

We did find however that although the mean amplitude of output signals was not repeatable within replicate measurements exposed to the same input frequency, the amount of variation observed between groups was quite consistent. This highlights therefore that the organisms were tolerant to the electrical input we administered and further, that measurement of amplitude would support the creation of analogue computing devices if a certain error tolerance were factored in.

We find that the transfer function for mycelia in the 900~Hz input treatment may be approximated efficiently with NARX modelling, the basis function for which was determined quantitatively to be a first order polynomial with 5 terms. This does not imply that the system under observation was linear, but rather the dominant components that are useful in prediction of future values were. 

Models generated with SISO methodology performed well, but were not as performant when doing inference runs using time series that had not been trained on (validation). Analysis of time series such as Fig. \ref{fig-model}b illustrate how the model struggles to generalise between different organisms (overfitting). This highlights how variable signal transformations may be between mycelia and the challenges associated with making repeatable measurements of living systems.

\section {Conclusions}
The majority of samples we tested transferred a signal of a frequency comparable to the input signal, indicating that mycelia can be used for electrical connection numerous parts of fungi-based organic electronic devices. Samples did not have a recoverable frequency when either no signal was broadcast or the input signal was modified enough to not be a harmonic. The former count occurs when no electrical connection was made with the organism, or when the electrode's surface area in contact with mycelium was inadequate to conduct a signal. 

Although the mean amplitude of output signals was not repeatable within repeated samples exposed to the same input frequency, the amount of variation detected between groups was relatively consistent. This shows that the organisms were tolerant to our electrical input and that measuring amplitude may allow the building of analogue computing systems with a certain error tolerance. By extension, our work demonstrates live systems' inherent variability and the need for practical unconventional computing device design to offset these effects.

Specifically, through generating a characteristic equation of signal transfer for a given mycelium, we find that it will be possible to derive characteristics of the input signal. It is therefore possible to encode via the medium of a mycelium-computer interface, which provides a basis for information processing. Future work should examine derivation of further forms of information, such as amplitude and conduction velocity.

Although no general model for mycelial transfer functions was uncovered, we regardless find that simulations may be generated reasonably efficiently with NARX modelling, which opens possibilities for creating simulated fungal computers through computationally tractable methods. 

\section{Authors' contributions} 
Conceptualisation, AA; measurements, NR, RM, NP; writing original draft, NR, NP; data analysis, RM; modelling, RM; writing, reviewing and editing, RM, NP, AA. All authors have read and agreed to the published version of the manuscript. All authors read and approved the final manuscript.

\section{Competing interests} 
All authors declare no competing interests.

\section{Acknowledgements}
We are grateful to Mayne Bio Analytics Ltd for support with data analysis and National Instruments Ltd for guidance on digital converter NI-USB-6210.

\section{Funding} 
This project has received funding from the European Union’s Horizon 2020 research and innovation programme FET OPEN “Challenging current thinking” under grant agreement No 858132.  The funders played no role in the design of the study and collection, analysis, and interpretation of data.

\section{Availability of data} 
The raw measurement dataset required to reproduce these findings is available to download from \url{https://zenodo.org/record/4430968#.Y8MJgOLP3z8}.


\end{document}